\documentclass[a4paper,fleqn,usenatbib]{pasj01}

\begin{document}
\Received{}
\Accepted{}

\title{Stellar wind accretion and accretion disk formation: applications to neutron star high mass X-ray binaries}

\author{Shigeyuki \textsc{Karino}\altaffilmark{1}%
}
\altaffiltext{1}{Fuculty of Science and Engineering, Kyushu Sangyo University, 2-3-1 Matsukadai, Higashi-ku, Fukuoka 813-8503, Japan}
\email{karino@ip.kyusan-u.ac.jp}

\author{Kenji \textsc{Nakamura},\altaffilmark{2}}
\altaffiltext{2}{Department of Mechanical Engineering, Kyushu Sangyo University, 2-3-1 Matsukadai, Higashi-ku, Fukuoka 813-8503, Japan}
\email{nakamura@ip.kyusan-u.ac.jp}

\author{Ali \textsc{Taani}\altaffilmark{3}}
\altaffiltext{3}{
Physics Department, Faculty of Science, Al-Balqa Applied University, 19117 Salt, Jordan
}
\email{ali.taani@bau.edu.jo}


\maketitle

\begin{abstract}

Recent X-ray observations have revealed the complexity and diversity of high-mass 
X-ray binaries (HMXBs). 
This diversity challenges a classical understanding of the accretion process onto the compact objects.
In this study, we reinforce the conventional concept of the nature of wind-fed accretion onto a neutron star 
considering the geometrical effect of radiatively accelerated wind, and re-evaluate the transported 
angular momentum by using a simple wind model. 
Our results suggest that even in an OB-type HMXB fed by stellar wind, a large amount of angular momentum 
could be transported to form an accretion disk due to the wind-inhomogeneity, if the binary separation is 
tight enough and/or stellar wind is slow.
We apply our model into actual systems such as LMC X-4 and OAO 1657-415, and discuss the possibility 
of disk formations in these systems.

\end{abstract}


\section{Introduction}

X-ray binary systems involving neutron stars are classified into two classes, according to their donor mass:
high-mass X-ray binaries (HMXBs) and low-mass X-ray binaries (LMXBs).
Among them, HMXBs are further classified into OB-type and Be-type \citep{C84,C86,BH91,B97,Pf02} according to
the type of the donor.
In general, in an OB-type system a neutron star captures quasi-spherical wind matter ejected
by a massive donor.
In this case, the accretion geometry around the neutron star takes almost spherical Bondi flow
\citep{HL39,BH44}.
If the accreting compact object is a {\it{magnetized}} neutron star, accreted matter would be captured
by the magnetosphere at a certain radius.
It is generally considered that the quasi spherical accretion flow could be trapped by strong magnetic field of
a neutron star before a formation of an accretion disk in an OB-type HMXB. 
With the recent growth of number of observed systems, however, the diversity of the accretion mode in HMXBs
has been revealed and peculiar systems which do not follow the traditional understanding have been found \citep{W15}.
For instance, the accretion modes in supergiant fast X-ray transients (SFXTs) are still under active discussions:
several theories have been verified such as clumpy wind accretion, magnetic gating mechanism due to strong field,
and/or settling spherical accretion shell \citep{Z05,WZ07,BFS08,S14}.
Furthermore, recently, discussions about the accretion and emission mechanisms of pulsating ultra-luminous X-ray
sources involving neutron stars has been started \citep{B14,E15,F16,I17a,I17b}.

\begin{figure}
 \begin{center}
	\includegraphics[width=6cm]{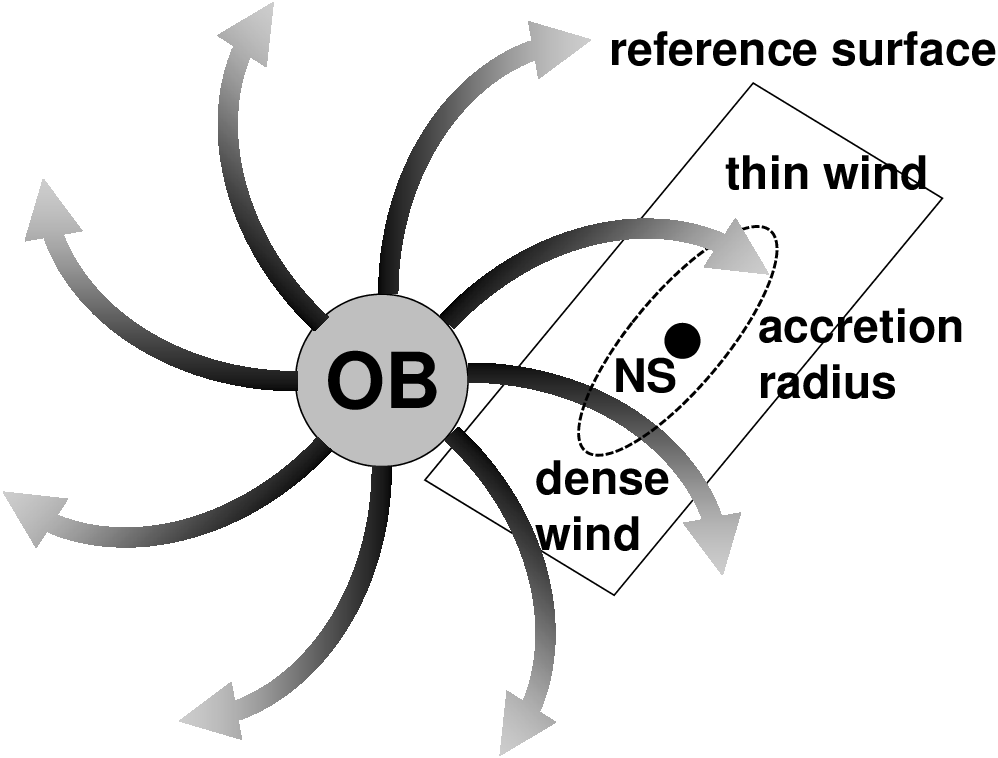}
 \end{center}
\caption{
Considered situation in this study.
The density of radiatively accelerated wind rapidly decreases once it is departed from the donor surface.
Because of the Coriolis force, the wind could be trailing and hence the accretion matter could have inhomogeneity.
}
\label{fig:concept}
\end{figure}

Although the study of OB-type HMXBs which are persistent bright X-ray sources has long history,
nature of OB-type HMXBs have not yet been understood completely.
OB-type systems have typically orbital periods of several days, and their X-ray luminosities are relatively persistent.
They occupy the upper-left region of Corbet diagram; their spin periods are systematically long even their orbits
are rather shorter than Be-type HMXBs \citep{C84,C86}.
It means that the transportation rate of the angular momentum via wind accretion is much lower than disk accretion.
In general, the spherical Bondi accretion is assumed in the wind-fed X-ray binaries and/or axisymmetric accretion is
often considered \citep{HL39, FR97, Edg04}. 
This simplification, however, cannot be valid in certain situations.
Especially, in tight binary systems with slow wind, an effect due to Coriolis force cannot be negligible \citep{HE13}.
In such systems, the relative velocity vector of the stellar wind becomes inclined to the binary axis. 
Furthermore, if the donor is massive star, the wind matter could be accelerated via the line-driven mechanism
\citep{CAK75}, and has steep gradient of the velocity and density in radial direction.
Due to this inclined accelerated wind, the accretion flow captured by a neutron star could have significant
inhomogeneities of density and velocity, as shown in Fig.~\ref{fig:concept}.
These inhomogeneities bring a certain amount of angular momentum onto the accreting neutron star,
and if it is enough large, an accretion disk may be formed around the neutron star.
Here, we examine the transferred angular momentum due to the inhomogeneity of the stellar wind
in tight binary systems and investigate the possibility of the disk formation in wind-fed X-ray binary systems.

In this study, we show that an accretion disk could be formed around the magnetized neutron
star even in a wind-fed binary system, when the binary separation is narrow and/or wind velocity is slow.
This result will play an important role to understand some peculiarities and diversities of observed HMXBs.
We apply our analysis of inhomogeneous wind accretion to the actual observed sources as following.
For example, LMC X-4, that is a well-known OB-type HMXB, has an accretion disk around the neutron star and
often shows large X-ray flares exceeding the Eddington luminosity \citep{I84,L91,MEW03}.
Additionally, OAO1657-415 shows peculiar binary parameters which cannot be explained by any binary evolution
scenario \citep{M12}.
The nature of these systems could be somehow different from the typical figure of OB-type HMXBs fed
via stellar wind \citep{T18a,T18b}.
In past, these systems have been considered to be fed via Roche lobe over-flow (RLOF) accretion \citep{F02}.
According to recent studies, however, it is suggested that the donors of these systems are rather small and cannot
fulfill their Roche lobe \citep{vdM07,R11,F15}.
In addition to this, it is needed to discuss the stability of the binary system that RLOF mass transfer proceeds from
a massive donor to relatively less-massive neutron star \citep{E06}.
In order to understand the nature of these peculiar systems, we need to revisit the nature of wind-fed accretion
from massive donors.
Therefore, we apply our model to these systems, and show that neutron stars in these systems could receive enough
amount of angular momentum from the stellar wind to form accretion disks.

In the next section, we briefly show the method of our analysis.
In Section 3, we show the results.
In Section 4, we discuss the validity and limits of our analysis.
Then we show some results of further applications to observed systems.
The final section is devoted to the conclusion.

\section{Transportation of angular momentum via stellar wind}

Transportation processes of angular momentum via stellar wind in X-ray binaries have been studied not only
by analytic approaches but also by numerical approaches.
For example, in \citet{SL76}, the authors suggested that a black hole fed by asymmetric stellar wind can form
an accretion disk when the orbital motion is taken into account.
On the other hand, they showed that a neutron star cannot have an accretion disk since the inner radius of 
the disk is too small.
Their work had been extended to investigate the spin-up/down evolution of the compact accreting object fed
by stellar wind by \citet{W81}.
Angular momentum transportation via non-uniform stellar wind has been studied numerically by many authors,
and especially after `90s, where the numerical simulations of wind accretion onto compact objects have been 
actively studied \citep{TJ93,RA95,R99,BP09}.
After the notable findings of the flip-flop instability in the supersonic accretion flows \citep{M87,FT88},
systematic studies have been done
\citep{M91,R97,R99}.
Highly resolved wind computations have revealed that, under certain conditions, angular momentum of 
the asymmetric wind could be transported to the circumference of the accreting object, and an accretion disk 
could be formed \citep{BP09,B13}.
Furthermore, when the wind is dense and slow, such as the slow wind emanating from giant/AGB donors, 
it is suggested that a static accretion disk could be formed without the flip-flop behavior \citep{TJ93,S04,HE13}.
Several such hydrodynamical studies of the interaction between giant donor and compact accretor have been
performed covering a wide range of phenomena \citep{J05,HC12,L17,EC17}.
In these works, however, they have not supposed a {\it{magnetized}} neutron stars as an accreting object.

On the other hand, in the context of X-ray binaries with OB-type donors, (quasi-)spherical wind accretion 
is considered in most cases \citep{S12,P17}; it is considered that the angular momentum is rather {\it withdrawn} 
from the neutron star due to the interaction between the neutron stellar field and the wind matter 
\citep{DP81,BFS08,G16}.
However, in most of previous studies, the line-acceleration of the wind from the OB-type donor \citep{CAK75}
is omitted, except for some simple analyses \citep{W81}.
The rapid acceleration produces a steep gradient of the wind velocity and density in radial direction.
When the orbital velocity is comparable to the wind velocity, the relative velocity vector of the wind 
could be inclined significantly (see Fig.~\ref{fig:concept}); and this radial gradient of the wind velocity could 
remarkably eliminate the symmetry of the wind to the accreting object \citep{HE13}.

Hence in this study, we consider wind accretion processes and the consequent angular momentum transport 
taking  such an asymmetry of the wind into account, using the simple wind model.
When the orbital motion is comparable to the wind velocity, the stellar wind is no longer symmetric 
about the binary axis.
Additionally, if the stellar wind suffered from line-driven acceleration, the wind density decreases rapidly.
As a result of this acceleration, the velocity and density distribution of the stellar wind passing 
the neutron star neighborhood becomes further asymmetric. 
These asymmetries bring some amount of angular momentum to the accreting neutron star.
In this study, we aim to evaluate the angular momentum transported due to this asymmetry.

In this purpose, we need an orbital radius $R_{\rm{orb}}$, and we obtain this from the Keplerian low;
\begin{equation}
R_{\rm{orb}} = \left[ \frac{G (M_{\rm{d}}+M_{\rm{NS}}) P_{\rm{orb}}^2}{4 \pi^2} \right]^{1/3} .
\end{equation}
Here, we assume that the donor mass is much larger than that of the neutron star ($M_{\rm{d}} \gg M_{\rm{NS}}$)
and the system takes a circular orbit.
In addition, the orbital velocity of the neutron star, $\vec{v}_{\rm{orb}} = v_{\rm{orb}} \vec{e}_{\rm{T}}$ can be
obtained by
\begin{equation}
v_{\rm{orb}} = \sqrt{ \frac{GM_{d}}{R_{\rm{orb}}} }.
\label{eq:vorb}
\end{equation}
$\vec{e}_{\rm{T}}$ denotes a unit vector in the direction of the orbital motion of the neutron star.
At the neutron star position ($R_{\rm{orb}}$), the wind velocity accelerated via line force
$\vec{v}_{\rm{w}} = v_{\rm{w}} \vec{e}_{\rm{R}}$ is computed by
\begin{equation}
v_{\rm{w}} = v_{\rm{inf}} \left( 1 - \frac{R_{\rm{d}}}{R_{\rm{orb}}} \right) ^{\beta} ,
\label{eq:vw}
\end{equation}
where $R_{\rm{d}}$ denotes the radius of the donor \citep{CAK75,KP00}.
$\vec{e}_{\rm{R}}$ denotes a unit vector in the radial direction seen from the center of the donor (see Appendix).
The acceleration parameter $\beta$ is fixed as $\beta = 1$.
$v_{\rm{inf}}$ denotes the wind velocity at the infinity.
From Eqs.~(\ref{eq:vorb}) and (\ref{eq:vw}), the relative velocity between the neutron star and the wind
can be obtained as following;
\begin{equation}
v_{\rm{rel}}^{2} = v_{\rm{w}}^2 + v_{\rm{orb}}^2 .
\end{equation}

Here we introduce a reference surface $S$, including the neutron star to evaluate the accretion rate of mass 
and angular momentum.
We set this surface to have its normal vector coincides with the relative velocity vector of the wind,
$\vec{v}_{\rm{rel}}$, at the neutron star position.

\begin{figure}
\begin{center}
\includegraphics[width=6cm]{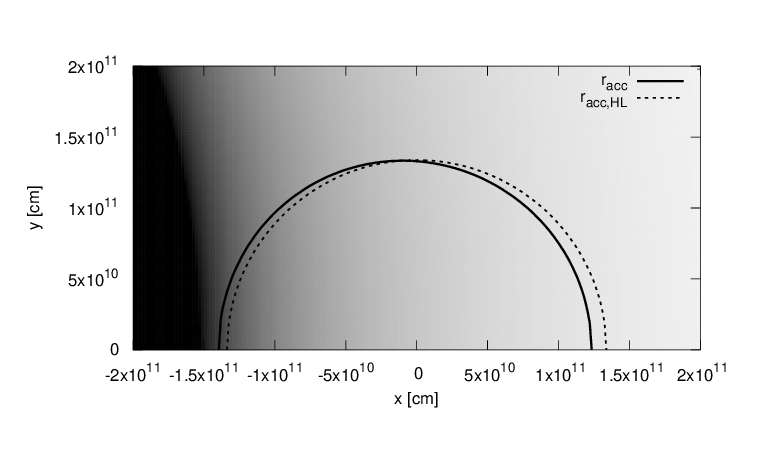}
\end{center}
\caption{Oblate accretion area defined by the balance between wind kinetic energy and the neutron star potential 
is shown by solid curve. 
At the same time, the accretion area evaluated from the wind velocity at the neutron star position 
(hitherto known Hoyle-Littleton theory) is shown by dashed curve. Density of the wind matter is also shown 
by gray scale: dark gray region corresponds to thick wind density.
The entire range from the pale end to the dark end covers 1 order magnitude of the wind density. }
\label{fig:Racc}
\end{figure}

The gravitational potential of the neutron star at each position on the $S$ surface is
\begin{equation}
U = - \frac{G M_{\rm{NS}}}{r} ,
\end{equation}
where $r$ is the distance from the neutron star on the $r$ surface.
If this potential overcomes the kinetic energy of the wind $K = v_{\rm{rel}}^2 / 2$, the wind matter passing
this point will be trapped by the neutron star.
Using an analogy from Hoyle-Littleton accretion theory \citep{HL39,FR97},
the area where the accretion matter would be trapped in the reference surface $S$ can be shown by
the maximum radius of such an area, $r_{\rm{acc}}$.
The shape of $r_{\rm{acc}}$ gets distorted depending on the wind and orbital parameters, as shown in 
Fig.~\ref{fig:Racc}.
In this figure, the locus of $U + K = 0$ on the $S$ surface is shown by the solid curve, while an accretion radius 
evaluated by the wind velocity at the neutron star position is shown by the dashed line (Hoyle-Littleton theory).
At the same time, the wind density on the $S$ surface is indicated by the gray value.
Clearly, the accretion area defined by $U + K = 0$ shifts toward the dense side.
Note that such a deformation effect of accretion region has been neglected in previous theoretical works
\citep{HL39, W81, FR97}.

The accretion rate of mass and angular momentum given by the accretion matter passing the area 
$r < r (U + K = 0)$ are computed as
\begin{equation}
\dot{M} = \int_{r<r (U+K=0)} \rho_{\rm{w}} v_{\rm{rel,z}} dS
\label{eq:Mdot}
\end{equation}
and
\begin{equation}
\dot{J} = \int_{r<r (U+K=0)} \rho_{\rm{w}} v_{\rm{rel,z}}^2 x dS ,
\label{eq:Jdot}
\end{equation}
respectively.
Here, $x$ is the projected length measured from the donor onto the orbital plane.
$v_{\rm{rel,z}}$ is the normal component of the relative velocity vector to the $S$ surface, and it is given 
in the procedure described in Appendix.

From Eqs.~(\ref{eq:Mdot}) and (\ref{eq:Jdot}), the specific angular momentum of the accretion flow can be 
computed as
\begin{equation}
\ell = \dot{J} \dot{M}^{-1} .
\end{equation}

\section{Results}

Next, we examine whether the accretion flow with the angular momentum obtained by our model can form
an accretion disk.
The accretion flow with its angular momentum $\ell$ will be circularized at the so-called circularization radius
$r_{\rm{circ}}$ that is defined as
\begin{equation}
r_{\rm{circ}} = \frac{\ell^2}{GM_{\rm{NS}}} ,
\end{equation}
and this radius corresponds to the initial size of the accretion disk \citep{F02}.
If the accretion flow is captured by the magnetic field out of this radius, however, the flow cannot form
an accretion disk and will be fall along the field line toward the polar regions of the neutron star \citep{BS76,F02}.
Hence, to form an accretion disk, the above circularization radius should be larger than the radius which
the magnetic force becomes dominant \citep{F02}:
\begin{equation}
r_{\rm{m}} = \left( \frac{B_{\rm{NS}}^4 r_{\rm{NS}}^{12}}{8 \xi^2 G M_{\rm{NS}} \dot{M}^2} \right)^{1/7} .
\end{equation}
This radius is called as a magnetospheric radius.
Here, $\xi$ denotes the accretion efficiency, and this value is simply assumed to be 0.5, hereafter.
At the same time, even if the circularization radius is larger than the magnetospheric radius, the mass accretion
could be inhibited by the centrifugal force; this situation is known as the propeller phase \citep{IS75,S86}.
This condition is written as $r_{\rm{m}} > r_{\rm{co}}$, where the corotation radius $r_{\rm{co}}$ is
\begin{equation}
r_{\rm{co}} = \left( \frac{G M_{\rm{NS}} P_{\rm{spin}}^2}{4 \pi^2} \right)^{1/3} .
\end{equation}
This inhibition of the accretion will be maintained until the neutron star loses enough angular momentum
and $r_{\rm{co}}$ becomes larger \citep{S86,BFS08}. 
As the conclusion, the condition that an accretion disk is formed in a wind-fed X-ray binary is
\begin{equation}
r_{\rm{m}} < r_{\rm{co}}, r_{\rm{circ}} .
\label{eq:condition}
\end{equation}
We investigate the condition where the above relation is settled, when we change various parameters.

In this study, we fix the mass, radius and magnetic field of the neutron star:
$M_{\rm{NS}} = 1.4 \rm{M}_{\odot}$, $r_{\rm{NS}} = 1.0 \times 10^6 \rm{cm}$, and
$B_{\rm{NS}} = 2.0 \times 10^{12} \rm{G}$, respectively.
Though the mass range of a neutron star is distributed between roughly $1.2 \rm{M}_{\odot}$ to
$\sim 2 \rm{M}_{\odot}$, the mass of the neutron star in HMXBs are not so far from  $1.4 \rm{M}_{\odot}$
\citep{SPR10,F15,Su18}.
Also the strength of the magnetic field of neutron stars in HMXBs are concentrated around a few of $10^{12} \rm{G}$
\citep{C16,T18a}.
The wind velocity is computed by Eq.~(\ref{eq:vw}). 
First, to grasp the general tendencies, we assume the mass of the donor and radius as
$M_{\rm{d}} = 15 \rm{M}_{\odot}$ and $R_{\rm{d}} = 10 \rm{M}_{\odot}$, respectively.
However, later on, we use rather realistic values when we make comparisons with the observed results.

\begin{figure}
\begin{center}
\includegraphics[width=6cm]{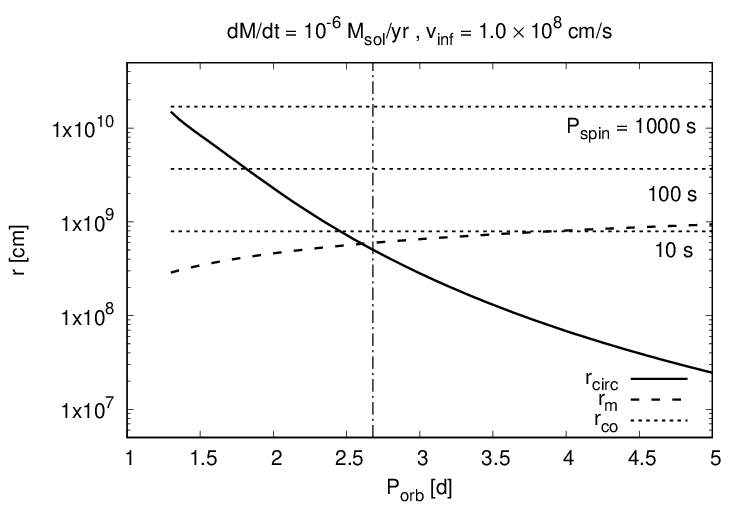}
\end{center}
\caption{Three radii, $r_{\rm{circ}}$, $r_{\rm{m}}$ and $r_{\rm{co}}$, are shown by solid, dashed and dotted lines, 
respectively, as functions of orbital period of the binary system.
The corotation radii are shown for three cases: $P_{\rm{spin}} = 10, 100$ and $1000 \rm{s}$.
The vertical line denotes the critical orbital period where mass accretion rate becomes 
$\dot{M}_{\rm{acc}} = 4 \times 10^{16} \rm{g \, s}^{-1}$.
Only in the left-hand-side of this vertical line, angular momentum could be transported to the neighbor of 
the neutron star.
}
\label{fig:xxx}
\end{figure}

With these settings, we vary the spin period of the neutron star $P_{\rm{spin}}$, mass loss rate of the donor
$\dot{M}_{\rm{w}}$, terminal velocity of the wind $v_{\rm{inf}}$, and the orbital period of the system $P_{\rm{orb}}$,
respectively.
As the typical result, in Fig.~\ref{fig:xxx} we show the computed radii ($r_{\rm{circ}}$, $r_{\rm{m}}$ and $r_{\rm{co}}$)
under the parameter setting as following:
$\dot{M}_{\rm{w}} = 1.0 \times 10^{-6} \rm{M}_{\odot} \rm{yr}^{-1}$, $v_{\rm{inf}} = 1.0 \times 10^{8} \rm{cm \, s}^{-1}$.
In this case, the maximum orbital period that relationship of Eq.~(\ref{eq:condition}) is satisfied is $2.6 \, \rm{d}$ 
regardless of the spin period.
We name this maximum orbital period where the condition of relation Eq.~(\ref{eq:condition}) is satisfied as 
$P_{\rm{orb,max}}$.
In the present case, the lines of propeller limits are well above of the magnetospheric radius in the region where
$P_{\rm{orb}} < P_{\rm{orb,max}}$ is satisfied.
In a system with rapidly rotating neutron star, however, the disk formation condition could be limited by
the propeller mass ejection (though fast spinning neutron stars have not been found in wind-fed HMXBs).

Even if the condition in Eq.~(\ref{eq:condition}) is satisfied, the subsonic accretion shell might rearranged the 
angular momentum before the disk is formed, when the accretion rate is too small \citep{S12}.
This critical mass accretion rate is given as 
\begin{equation}
\dot{M}_{\rm{acc}} > 4 \times 10^{16} \rm{g \, s}^{-1} .
\label{eq:condition2}
\end{equation}
This limiting accretion rate for the shell forming is shown by vertical dashed-dotted line in Fig.~\ref{fig:xxx}.
Only in the left-hand side of this vertical line, the mass accretion rate is enough high and an accretion disk 
can be formed (see Sec.~\ref{sec:dis}).
In the case shown in Fig.~\ref{fig:xxx}, the conditions Eqs.~(\ref{eq:condition}) and (\ref{eq:condition2}) are 
satisfied when the orbital period is shorter than $2.6 \, \rm{d}$.

\begin{figure*}
\begin{center}
\includegraphics[width=5.5cm]{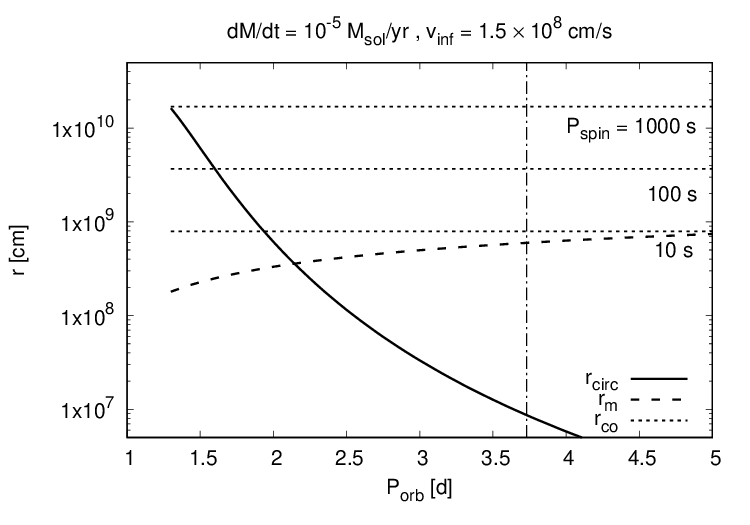}
\includegraphics[width=5.5cm]{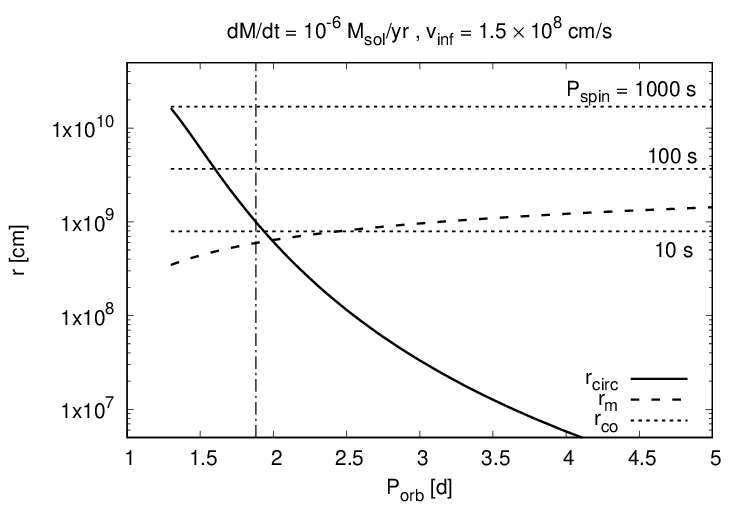}
\includegraphics[width=5.5cm]{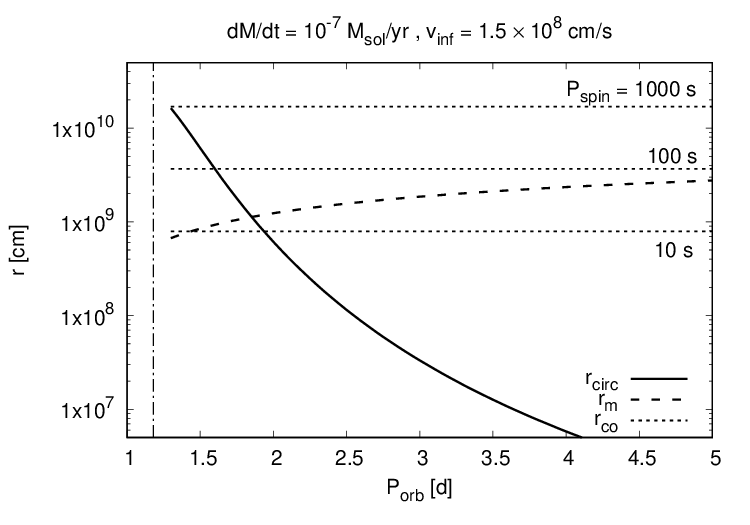} \\
\includegraphics[width=5.5cm]{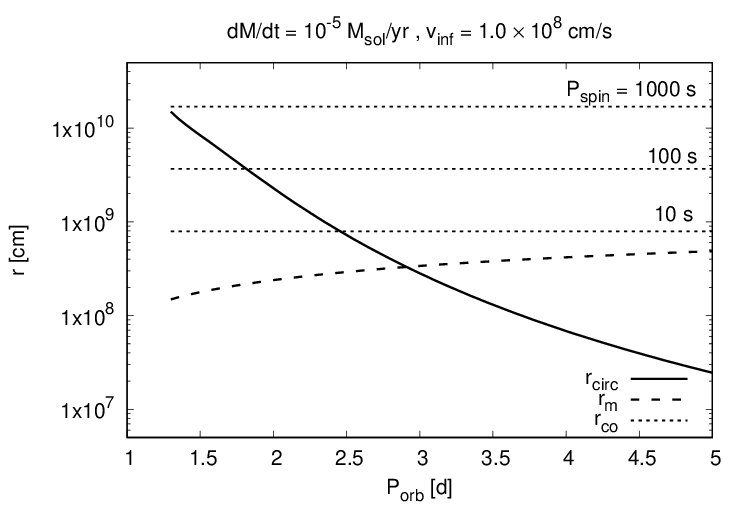}
\includegraphics[width=5.5cm]{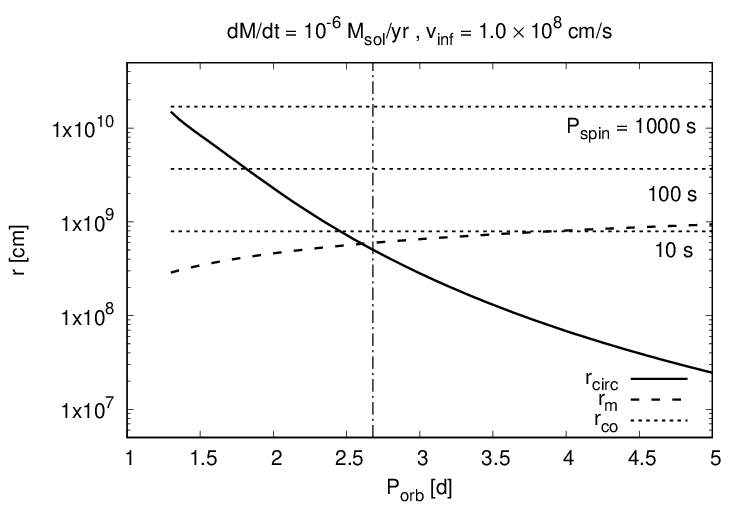}
\includegraphics[width=5.5cm]{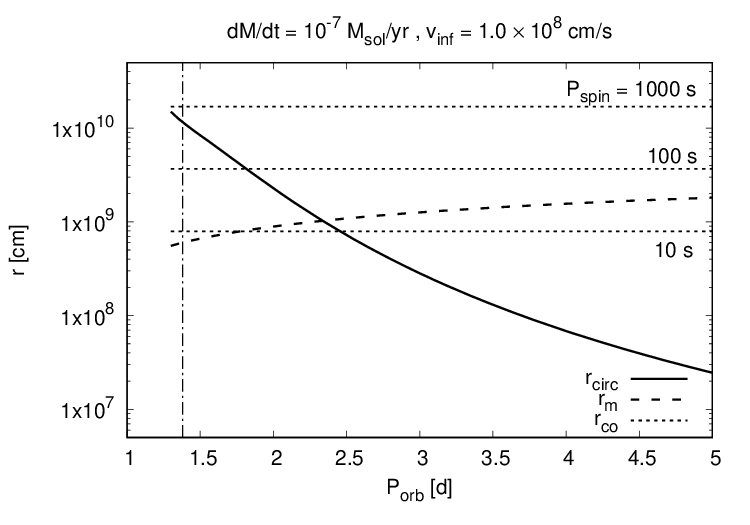} \\
\includegraphics[width=5.5cm]{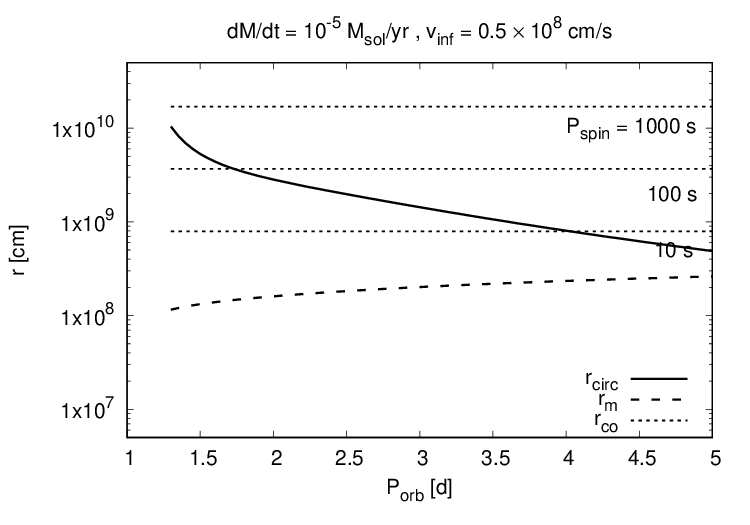}
\includegraphics[width=5.5cm]{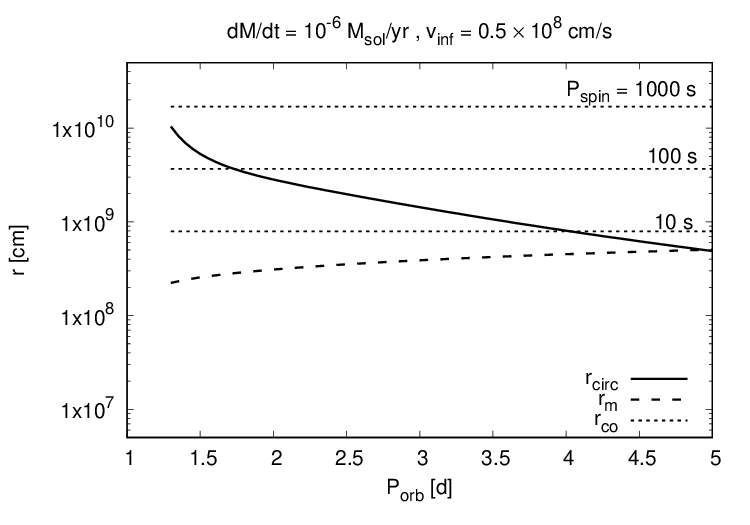}
\includegraphics[width=5.5cm]{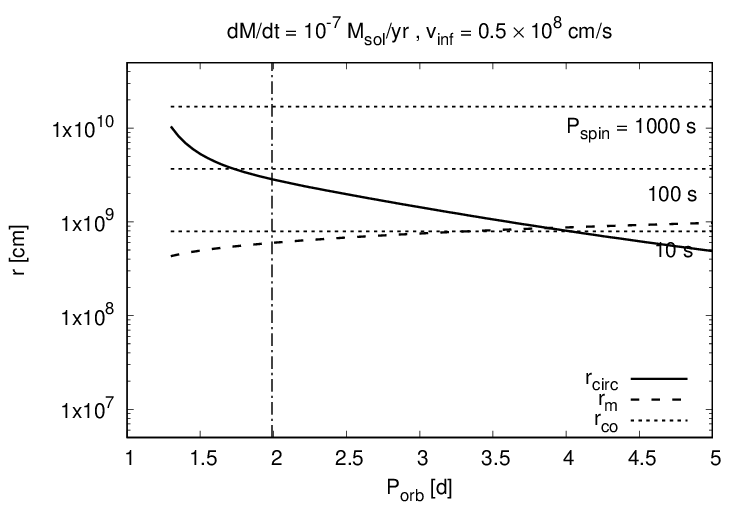} \\
\end{center}
\caption{The same figures with Fig.~\ref{fig:xxx} for different mass loss rate and wind velocity.
For the cases without vertical lines, the accretion rates are enough large in entire parameter space
in the figures.
}
\label{fig:all}
\end{figure*}

In Fig.~\ref{fig:all}, we show the results obtained by changing parameters.
Here, we vary the mass loss rate of the donor
($\dot{M}_{\rm{w}} = 1.0 \times 10^{-5}$, $1.0 \times 10^{-6}$ and $1.0 \times 10^{-7} \rm{M}_{\odot} \rm{yr}^{-1}$)
and the wind velocity ($v_{\rm{inf}} = 0.5, 1.0$ and $5.0 \times 10^{8} \rm{cm \, s}^{-1}$).
In each case, we examine three spin periods of the neutron star ($P_{\rm{spin}} = 10, 100$ and $1000 \rm{s}$).
When the wind velocity is slow ($v_{\rm{inf}} = 5 \times 10^{7} \rm{cm \, s}^{-1}$), the disk forming condition,
Eq.~(\ref{eq:condition}), is satisfied in broad range of $P_{\rm{orb}}$.
However, when the mass loss rate of the donor is small, subsonic shell forming condition, 
Eq.~(\ref{eq:condition2}), dominates the possibility of the disk formation.

\begin{figure}
\begin{center}
\includegraphics[width=6cm]{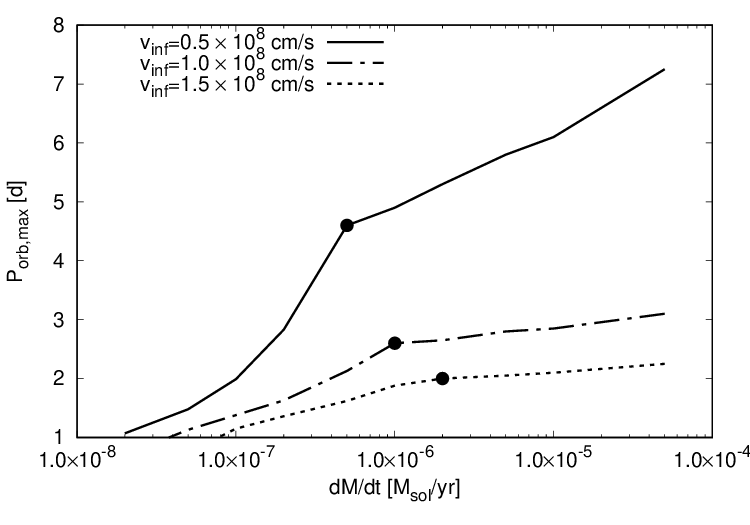}
\end{center}
\caption{The maximum orbital period to form an accretion disk, as functions of mass loss rate of the donor.
Results for three different $v_{\rm{inf}}$ are shown. 
The black dots indicate the exchange of the dominant mechanism to restrict the disk formation.}
\label{fig:MdotPorb}
\end{figure}

In Fig.~\ref{fig:MdotPorb}, we show the maximum orbital period to form an accretion disk, $P_{\rm{orb,max}}$,
as functions of mass accretion rate and wind velocity.
In this figure the mass loss rate is chosen as a variable parameter and the results are shown for three different
wind velocities.
From the figure, we can see that the wind velocity causes a significant effect on the disk formation.
When the wind velocity is enough large (say, $v_{\rm{inf}} >1.0^{8} \rm{cm \, s}^{-1}$), the dependence of 
$P_{\rm{orb,max}}$ on the mass loss rate is rather weak and the critical orbital period is quite short 
(less than $3 \, \rm{d}$ and shorter than typical period of HMXBs).
On the other hand, when the wind velocity is slow, the dependence on the mass loss rate becomes significant
and the maximum orbital period becomes much longer when the mass loss rate is large.
The change of the gradient around $\dot{M}_{\rm{w}} \approx 1 \times 10^{-6} \rm{M}_{\odot} \rm{yr}^{-1}$, 
in Fig.~\ref{fig:MdotPorb}, stems from a change of the dominating conditions; below this point the shell forming 
condition Eq.~(\ref{eq:condition2})
limits the disk formation.
On the other hand, when the mass loss rate is larger than this point, the disk formation is limited by the condition
Eq.~(\ref{eq:condition}).
These exchanges are shown with black dots in the same figure.

Possibility of a disk formation around a compact object fed by slow wind has been previously suggested 
\citep{W81,S12,EM18a,T18a}.
In \citet{DSP10}, they considered the disk forming possibility in SFXTs which are accreted by slow wind, 
to explain burst activities of SFXTs. 
In our treatment of deformed accretion region (Fig.~\ref{fig:Racc}), we have found that further angular momentum 
could be transported and the possibility of disk formation becomes significant (this will be discussed further 
in section 4).

We can summarize our results as following: 
\begin{itemize}
\item For systems with short orbital periods (for tight systems), the necessary condition for the disk formation
($r_{\rm{m}} < r_{\rm{circ}}$) could be satisfied.
\item When the spin of the neutron star is fast, however, accretion could be inhibited due to propeller effect.
For typical spin period of neutron stars in OB-HMXBs, however, it does not dominantly restrict the disk formation. 
\item When the wind velocity is slow, an accretion disk can be formed even in systems with large orbital period.
\item When the mass loss rate of the donor is large, an accretion disk can be formed even in systems with
large orbital period.
When the mass loss rate of the donor is small, however, the disk formation could be limited by the subsonic 
shell forming condition.
\end{itemize}

\section{Discussion}
\label{sec:dis}

\subsection{Deformation of the accretion region}

In this study, we have shown that even in wind accretion regime enough amount of angular momentum 
to form an accretion disk is transported.
Furthermore, we have shown that it is crucial to consider the orbital motion of the neutron star and 
the acceleration of the wind matter.
If we consider these two effects, the accretion region shifts inward of the orbit as shown in Fig.~\ref{fig:Racc}.
Because of this shift of accretion region, the neutron star can capture an extra dense wind matter from
the inner region.
And this makes possible to capture a large amount of angular momentum from the wind.
At the same time, it could be understood in Fig.~\ref{fig:Racc} that though the wind matter accreted from
the outer region has a large velocity, its density is much lower and the accretion region is rather narrower.
Namely, inward deformation of the accretion region is essentially important to estimate the accretion rate
of the angular momentum.

\begin{figure}
\begin{center}
\includegraphics[width=6cm]{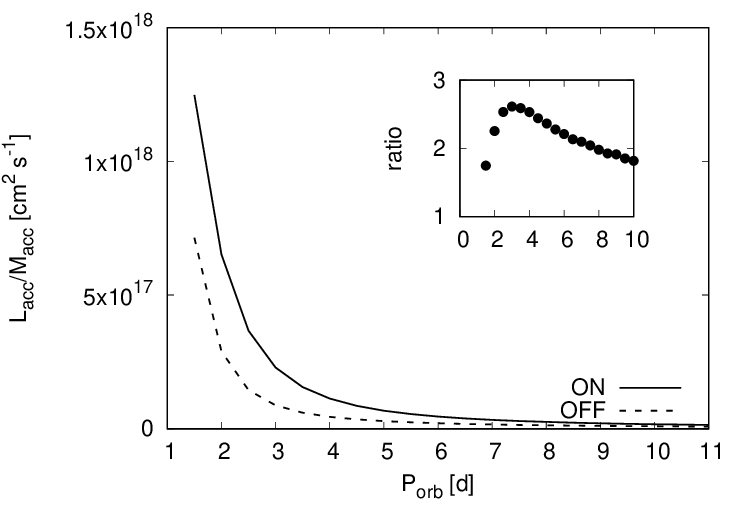}
\end{center}
\caption{The transported specific angular momentum, when the deformation of the accretion region is considered
(ON) and that without this effect (OFF).
In the small box, the ratio of this specific angular momentum ($\ell_{\rm{ON}} / \ell_{\rm{OFF}}$)as a function of 
the orbital period is shown.  
The parameters of the system is the same with the case shown in Fig.~\ref{fig:xxx}.
}
\label{fig:ratio}
\end{figure}

In order to depict the effect of the deformation of the accretion region clearly, an illustration of the result 
is given in Fig.~\ref{fig:ratio}, where the transported specific angular momentum is shown in two cases. 
For the first case the deformation of the accretion region is not considered.
This is the case which is shown by dashed curve in Fig.~\ref{fig:Racc} and we refer this case as OFF.
On the other hand, the result obtained with the consideration of the deformed accretion region is shown at the 
same time by solid curve. 
This second case is corresponding to the accretion region shown by solid line in Fig.~\ref{fig:Racc}.
We refer this second case as ON.
Clearly, in the case ON, the accreted specific angular momentum increases. 
Though this deformation effect has been considered less serious in previous works \citep{HL39,SL76,W81}, 
this result shows that the deformation could play rather important role in the nature of accretion mode 
in tight binary systems.

At the same time, in the small box of Fig.~\ref{fig:ratio}, we show the ratio of the specific angular momentum 
between these two cases as a function of the orbital period.  
In the present parameter set, this ratio takes maximum at $P_{\rm{orb}} = 3 \rm{d}$.
This behavior could be understood as follows.
When the orbital period is long (i.e. orbital separation is large), the wind from the donor has already well 
accelerated ($v_{\rm{w}} = v_{\rm{inf}} \gg v_{\rm{orb}}$).
Hence, the wind could be considered like a planar uniform flow.
Then the effect of the deformation of the accretion region could be smaller.
On the other hand, when the orbital separation is small, the orbital velocity dominates in the relative velocity.
Therefore, the acceleration of the wind does not have an important role.
As the conclusion, the deformation of the accretion region becomes important at the intermediate orbital period
around $3 \rm{d}$.

\subsection{Formation of accretion disk}

Fig.~\ref{fig:MdotPorb} indicates that an accretion disk could be formed even when the orbital period of 
the system is rather large, if the wind velocity is slow.  
The slow wind velocity is caused by slow terminal velocity due to certain reasons such as line-driven efficiency 
and X-ray photo-dissociations \citep{DSP10,K12,K14}.
At the same time, wind velocity becomes slower also due to the increase of the relative radius of the donor to
the orbital radius as seen in Eq.~(\ref{eq:vw}).
In the above discussion, however, we have only shown that a certain amount of angular momentum could be
transported inside the accretion radius.
Deep inside of the accretion radius, it is required to consider whether an accretion disk could really be formed or not.

In general, the captured matter from the stellar wind forms a shock during its falling path to the neutron star.
If the cooling is ineffective, falling matter could form a quasi-static settling shell below the shock \citep{S12}.
Once such a shell is formed, the angular momentum is transported due to turbulence caused by
hydrodynamic instabilities inside the shell.
Hence, the transported angular momentum will be redistributed before it reaches deep inside of the accretion 
radius.
When the accretion rate is enough large, $\dot{M} > 4 \times 10^{16} \rm{g \, s}^{-1}$, however,
a settling shell is not formed and accretion matter could be fallen onto the neutron star quickly.
This limiting accretion rate for the shell forming is shown by vertical dashed-dotted line
in Figs.~\ref{fig:xxx} and \ref{fig:all}.
This line shows the limiting case where the mass accretion rate evaluated by Eq.~(\ref{eq:Mdot}) becomes
the critical value: $4 \times 10^{16} \rm{g \, s}^{-1}$.
In the right-hand side of this vertical line, a quasi-spherical subsonic shell will be formed and in the shell
the accreted angular momentum will be redistributed.
In this case, therefore axisymmetry recovers and the accretion disk cannot be formed.
On the other hand, in the left-hand side of the vertical line, the mass accretion rate is enough high
and a settling shell cannot be formed.
Then the angular momentum will be transported to deep inside of the accretion region.
Finally, if the condition given by Eq.~(\ref{eq:condition}) is satisfied, an accretion disk would be formed.

The above condition onto the mass accretion rate is also important to consider the nature of the formed
accretion disk.  
When the mass supply is small, the accretion disk would 
be a radiatively inefficient accretion flow (RIAF), and its X-ray luminosity steeply decreases with mass 
accretion rate \citep{NY94,NY95,KMF08}.
The accretion disk will be RIAF-like only when the accretion rate is rather less and the temperature is high. 
Hence, if our disk formation condition is satisfied (high accretion rate), it seems reasonable to form a disk 
which behaves as a standard disk \citep{SS73}.
We hope future observations with more sensitive techniques can test this possibility.

To fix the position of the inner edge of the accretion disk, the effect of magnetic field is crucially important.
Moreover, the neutron star magnetosphere plays an important role to onset the propeller mass ejection
\citep{S86,BFS08}.
In this study, we assume a constant magnetic field strength; $B = 2 \times 10^{12} \rm{G}$.
Actually, the neutron star magnetic field could take from the lower limit ($\sim 10^9 \rm{G}$) seen in
LMXB systems to the higher end ($\sim 10^{14} \rm{G}$) observed in magnetars \citep{Pf02,P04,vdH09}:
it spans 5 orders of possibilities.
According to the straightforward observations of cyclotron resonance features in HMXB systems, 
however, most of neutron stars in HMXBs show intermediate level of the field strength, 
$\sim 10^{12} \rm{G}$ \citep{T18a,T18b}.
Although this stems to a certain observational bias, it is worth mentioning that the magnetic fields are still
$\sim 10^{13} \rm{G}$ even in most active neutron X-ray sources such as LMC X-4 and NGC300 ULX-1
\citep{ME01,MEW03,LRZ00,C18}.
Therefore our assumed magnetic field $2 \times 10^{12} \rm{G}$ is not an odd choice as the first attempt.

Considering a system which orbital period is slightly shorter than the critical period, $P_{\rm{orb,max}}$,
in such a system, the difference between two radii ($r_{\rm{m}}$ and $r_{\rm{circ}}$) are small, and the size
of formed disk may be very small.
Once an accretion disk is formed, however, due to viscous effects the angular momentum of the disk matter
could be transported from inside towards the outer edge.
At the same time the matter near the outer edge would be pushed out and disk itself would expand.
At the expanded disk edge, subsequent accreted matter will be attached and the size of the disk would be
further extended.
For accrual example, wind-fed HMXB system LMC X-4 has a warped accretion disk enough large to shade
the X-ray from the neutron star, even though it is expected that $r_{\rm{circ}} \approx r_{\rm{m}}$ and 
the disk size could be small (discussed later).

In observed systems, the orbital eccentricities are not exactly zero.
The eccentricity changes the geometrical relation of the neutron star to the stellar wind and affects
the results.
In typical Be-type HMXBs, the orbital eccentricity is rather large; it is considered to be induced by
the natal kick when the neutron star was born in core collapse supernova.
On the other hand, however, a OB-type HMXB typically has small ($e <0.1$) eccentricity \citep{B97}.
This is because of strong circularizing effect caused by the interaction between the neutron star
and dense wind matter.
The circularizing time could be much smaller than the HMXB life-time.
Hence, in the present situations, we could neglect the effect of such orbital eccentricities.

\subsection{Diversity of the donor and neutron star}

\begin{figure}
\begin{center}
\includegraphics[width=6cm]{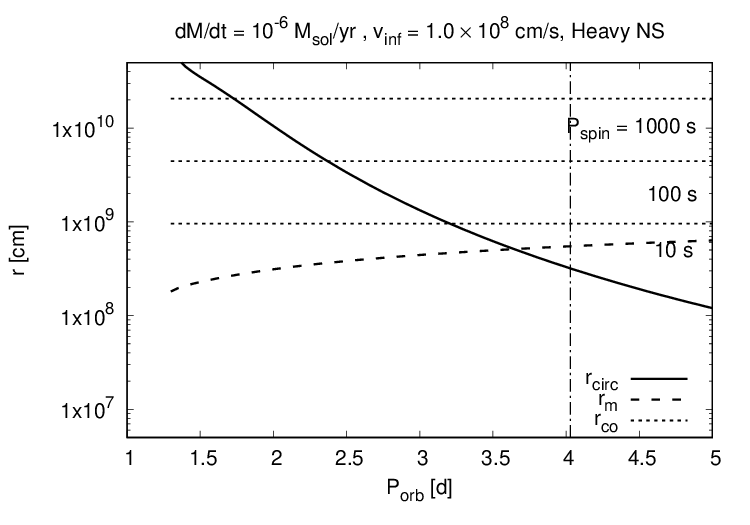}
\end{center}
\caption{The same figure with Fig.~\ref{fig:xxx} but for a heavy NS: $M_{\rm{NS}} = 2.5 \rm{M}_{\odot}$.
The other parameters are the same with Fig.~\ref{fig:xxx}.
The maximum orbital period, $P_{\rm{orb,max}}$ is extended up to 3.5 d.
}
\label{fig:heavyNS}
\end{figure}

In this study, we have considered only a typical neutron star whose mass is $1.4 \rm{M}_{\odot}$.
Recent studies have availed, however, that some neutron stars have larger mass, up to $2 \rm{M}_{\odot}$
\citep{vK95,R11,F15}.
Such heavy neutron stars made strong constraint onto the equation of state of the neutron stellar matter
and play important role to understand the nature of condensed matter.
In the present study, such a large mass of neutron star also changes the geometry of the accretion region.
Deep potential broadens the accretion radius especially in slow wind side and enhances asymmetry of
wind geometry.
This asymmetric wind brings further extra angular momentum to the neutron star neighbor and relaxes
the condition of disk formation.
In order to evaluate this effect, we show the result for a heavy neutron star with
$M_{\rm{NS}} = 2.5 \rm{M}_{\odot}$ in Fig.~\ref{fig:heavyNS}.
In this example, the other conditions are the same with the case shown in Fig.~\ref{fig:xxx}.
The maximum orbital period, $P_{\rm{orb,max}}$ is extended up to $3.5 \rm{d}$, in this case.
Since the mass of neutron star cannot largely exceed $2 \rm{M}_{\odot}$, we can conclude that the effect
of neutron stellar mass could be limited in the present context.

\begin{figure}
\begin{center}
\includegraphics[width=6cm]{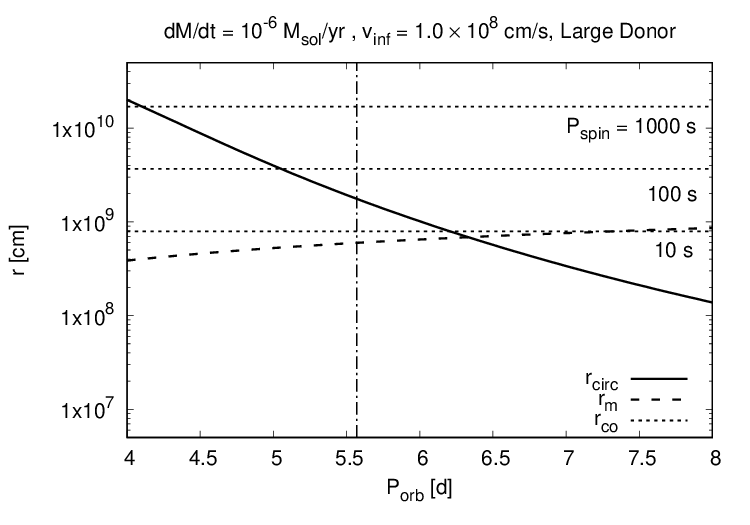}
\end{center}
\caption{The same figure with Fig.~\ref{fig:xxx} but for a large donor: $R_{\rm{d}} = 20 \rm{R}_{\odot}$.
The other parameters are the same with Fig.~\ref{fig:xxx}.
The maximum orbital period, $P_{\rm{orb,max}}$ is limited at 5.3 d due to subsonic shell formation.
}
\label{fig:largedonor}
\end{figure}

We have considered only a typical SG donor with $M_{\rm{d}} = 15 \rm{M}_{\odot}$ and
$R_{\rm{d}} = 10 \rm{R}_{\odot}$.
The wind velocity at the neutron star position, however, depends on the donor size (see, Eq.~(\ref{eq:vw})).
That is, when the donor size is large, the wind velocity becomes slower since the zone of acceleration
effectively decreases.
As the consequence, the possibility of disk-formation could be larger.
To confirm the effect of donor size, we have additionally computed the same case study for a large donor
with $R_{\rm{d}} = 20 \rm{R}_{\odot}$.
The result is shown in Fig.~\ref{fig:largedonor}.
In this case, the parameters except for the donor radius are the same with the case of Fig.~\ref{fig:xxx}.
We can see that, in this case, the condition Eq.~(\ref{eq:condition}) is satisfied in a broader range
$P_{\rm{orb}} < 6.4 \rm{d}$ for all spin periods.
This time, however, the limit of subsonic shell formation dominates the maximum orbital period for disk-formation.
As the result, the system which orbital period is less than $5.3 \rm{d}$ could form an accretion disk around
the neutron star. 
This limit will be pushed up due to high mass-loss rate of the donor (see Fig.~\ref{fig:all}).
Since a large evolved star tends to increase its mass-loss rate at the same time, an evolved donor would result
in a large possibility to form an accretion disk around the neutron star.

\subsection{Observed systems}

\subsubsection{LMC X-4}

LMC X-4 is one of the most powerful persistent OB-type X-ray binary in our neighborhood.
It emits luminous X-rays with $L_{\rm{X}} = 3 - 4 \times 10^{38} \rm{erg \, s}^{-1}$ persistently,
and often causes bright X-ray flares achieving $L_{\rm{X}} \sim 10^{40} \rm{erg \, s}^{-1}$
that exceeds the Eddington limit \citep{MEW03,S18}.
The spin period, orbital period and orbital eccentricity is 13.5 s, 1.4 d and $< 0.003$, respectively
\citep{L78,W78,K83}.
From the spectral properties, the donor star is identified as an O8III type super-giant star,
which mass is assumed to be $M_{\rm{d}} = 18 \rm{M}_{\odot}$ \citep{CI77,F15}.
Besides the orbital period, it shows 30.5 d periodicity and it is suggested to be caused by a warped
accretion disk around the neutron star \citep{I84,L91}.

In this bright system, since the existence of the accretion disk around the neutron star is evident,
it has been considered that the accretion matter is supplied via RLOF \citep{LRZ00,PBV02}.
Recent analysis, however, reveal that the radius of the donor remains confined to $7 - 8 \rm{R}_{\odot}$
\citep{R11,F15}.
Assuming that the neutron star mass is $1.4 \rm{M}_{\odot}$, the Roche lobe size is
much larger than the donor radius.
Hence, the donor could remain well inside the lobe and RLOF has not been started.
On the other hand, also it has been suggested the ejection of strong wind from the donor
\citep{V97,B99}.
Then the wind-fed accretion in this system is suspected.

\begin{figure}
\begin{center}
\includegraphics[width=6cm]{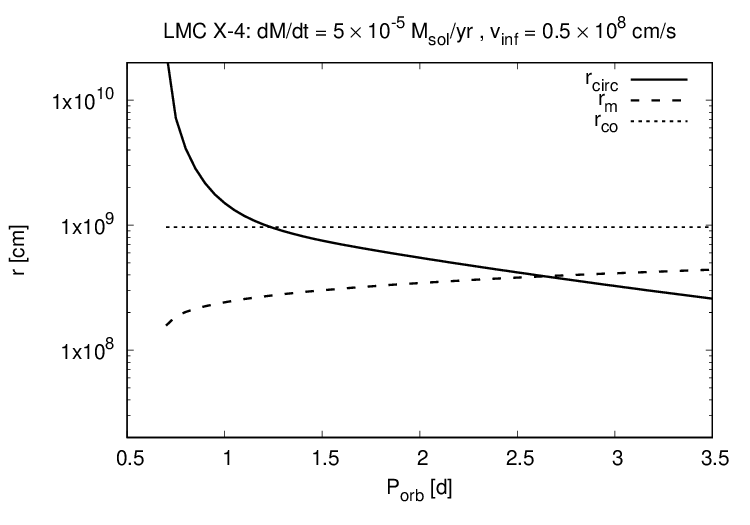}
\end{center}
\caption{This is the figure for LMC X-4.
The binary parameters are summarized in Table. 1.}
\label{fig:LMCX-4}
\end{figure}

Given these circumstances, we consider the possibility of the disk formation due to angular momentum
accretion via stellar wind in this system.
In this purpose, we use the same analysis described in the previous sections.
The used parameters are summarized in Table.~\ref{tab:first}.
The corresponding radii ($r_{\rm{circ}}$, $r_{\rm{m}}$ and $r_{\rm{co}}$) are shown in Fig.~\ref{fig:LMCX-4}.
From the condition, Eq.~(\ref{eq:condition}), in this system, disk formation via stellar wind could be possible
if the orbital period is shorter than $2.5 \rm{d}$.
Since the orbital period of this system is $1.4 \rm{d}$, and since the propeller effect could be avoided
with its orbital period, it is strongly suggested that the disk formation via wind accretion could be possible.

Additionally, in this system, the difference between 
$r_{\rm{circ}}$ and $r_{\rm{m}}$ is rather small. 
It means that the accretion regime (direct wind accretion / disk accretion) could be converted easily
if the situation varies.
Observation shows that this system fluctuates between the spin-up and spin-down regime 
\citep{M17}.
Such a spin evolution might be related to such a marginal situation. 
Since the exchange of accretion regime may change the spectral feature, multifaceted studies are highly
required \citep{Pa02,NP03}.

\subsubsection{OAO 1657-415}

OAO 1657-415 was identified in 1970's, and has been one of the most famous HMXBs.
The orbital period and orbital eccentricity is 10.448 d and 0.107, respectively \citep{C93}.
The donor star was identified as Ofpe star, however, since hydrogen line is weak, it might be
a Wolf-Rayet star, as like Cyg X-3 \citep{M12}.
By optical observations, the parameters of the donor have been measured as
$M_{\rm{d}} = 14.3 \rm{M}_{\odot}$,
$R_{\rm{d}} = 24.8 \rm{R}_{\odot}$, $\dot{M}_{\rm{w}} = 2 \times 10^{-6} \rm{M}_{\odot} \rm{yr}^{-1}$,
respectively \citep{M12}.
The wind velocity is suggested to be very slow ($v_{\rm{inf}} = 2.5 \times 10^7 \rm{cm \, s}^{-1}$),
and it may due to a displacement from spherical symmetry caused by a neutron star.
It is known that the standard binary evolution scenario cannot derive such a set of binary parameters
of this system \citep{M01,M12}.
Although no firm evidence of an accretion disk has been reported, the shape of the cumulative luminosity
distribution of this system is similar to disk-forming systems \citep{SP18}.

\begin{figure}
\begin{center}
\includegraphics[width=6cm]{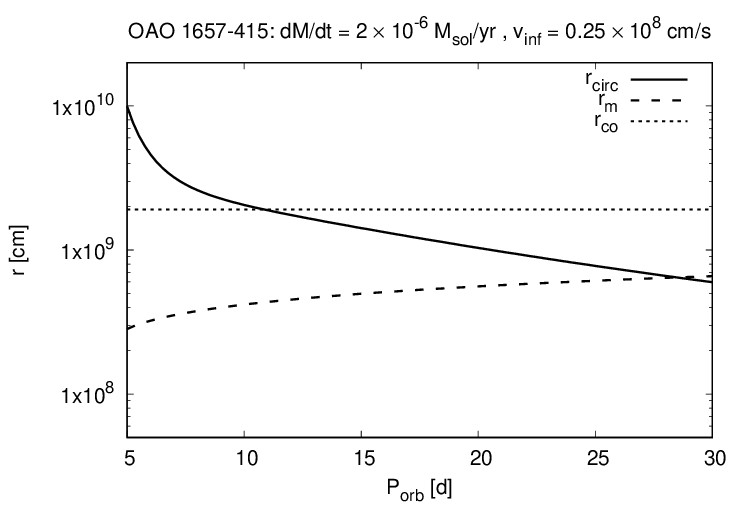}
\end{center}
\caption{This is the figure for OAO1657-415.
The binary parameters are summarized in Table. 1.}
\label{fig:OAO1657}
\end{figure}

We apply our model to this system with the same procedure that is discussed above.
(Since the radius of the donor and the size of the Roche lobe size is almost the same in this system,
it is a sensitive issue whether RLOF proceeds or not.)
The results of each radii computed with the parameters of OAO1657-415 is shown in Fig.~\ref{fig:OAO1657}.
From this figure, if the orbital period is shorter than 23 d, the disk formation could be possible.
In fact, the orbital period of this system is 10.4 d and it clearly satisfies this condition.
Also we can see that the propeller effect and subsonic shell forming do not work in this parameter range.
Therefore, we conclude that also in this system, the disk formation could be realized via wind-fed accretion.

In this system, the slow wind (due to slow wind and large donor size) plays an important role in the disk formation.
Such a disk formation in slow-wind binary is also suggested by numerical computations \citep{HE13}, and this
consistency supports the robustness of our working hypotheses.

Very recently it has been reported that a disk-like structure could be formed even when the donor 
does not fill its Roche lobe under certain conditions, according to the numerical computations
\citep{EM18a,EM18b}.
Such a marginal accretion mode (wind RLOF / beamed wind) might be a key to understand the diversity 
and complexity of HMXBs.

\begin{table}
  \tbl{Important parameters of HMXB systems: LMC X-4 and OAO1657-415}{%
  \begin{tabular}{lcc}
      \hline
      Name & LMC X-4 & OAO 1657-415 \\
      \hline
      $M_{\rm{d}} [\rm{M}_{\odot}]$ & 18 & 17.5 \\
      $R_{\rm{d}} [\rm{R}_{\odot}]$ & 7.7 & 25 \\
      $P_{\rm{spin}} [\rm{s}]$ & 13.5 & 37.7 \\
      $P_{\rm{orb}} [\rm{d}]$ & 1.4 & 10.4 \\
      $v_{\rm{inf}} [10^{8} \rm{cm \, s}^{-1}]$ & 0.5 & 0.25 \\
      $\dot{M}_{\rm{d}} [10^{-6} \rm{M}_{\odot} \rm{yr}^{-1}]$ & 50 & 2 \\
      $B_{\rm{NS}} [10^{12} \rm{G}]$ & 11.2 & 4 \\
      \hline
    \end{tabular}}\label{tab:first}
\begin{tabnote}
The binary parameters that we have used in the analysis.
References are in the text.
\end{tabnote}
\end{table}


\section{Conclusion}

In this study, we have revisited to the transportation rate of the angular momentum in the wind-fed
neutron star X-ray binaries.
We have taken into account the asymmetry due to the orbital motion of the neutron star and
the wind acceleration.
The asymmetry of the wind can lead to a deformation of the accretion region and due to this, 
a large amount of angular momentum could be transported inside the accretion radius even 
in wind accretion.
Furthermore, we have shown that an accretion disk could be formed around the neutron star
under certain situations;
when the wind velocity is slow because of a small terminal velocity and/or inefficient wind acceleration
due to a large donor radius, the possibility of disk formation could be larger.

We applied our disk formation criteria to the observed peculiar systems hitherto-considered
to be fed via RLOF.
In our analysis, neutron stars in LMC X-4 and OAO1657-415 could have accretion disks,
even if they were fed via stellar winds.
For the LMC X-4, the orbital period is short enough to capture plenty amount of angular momentum 
from the wind of the donor. 
In addition, the propeller effect could be avoided with this orbital period.
This results elucidates the source of a warped accretion disk in this object. 
While for the OAO1657-415, the slow wind enhances the transportation rate of the angular momentum 
and gains the possibility of the disk formation.

Finally, the existence of the accretion disk will affect the observed properties such as spectral hardness,
and the spin evolution of the accreting neutron star.
To reveal the accretion nature of HMXBs, further studies both in observations and theoretical
sides are required.

\begin{ack}

We thank the referee for fruitful suggestions.
This work was supported by JSPS KAKENHI Grant Number 18K03706.
Ali Taani gratefully acknowledges the Institute of High Energy Physics,
Chinese Academy of Sciences through CAS-PIFI  Fellowship 2018
\end{ack}

\appendix
\section*{Tips of computations}

Here, the coordinate system which we have used to evaluate the accretion rate of angular momentum
via line-driven stellar wind is described.
At the beginning, we define a reference surface $S$ of the wind accretion to evaluate the accreted
angular momentum.
We find the relative velocity vector of the wind at the position of the neutron star, $\vec{v}_{\rm{rel}}$.
The relative velocity can be obtained by a simple addition of the orbital velocity
$\vec{v}_{\rm{orb}} = v_{\rm{orb}} \vec{e}_{\rm{T}}$ (see, Eq.~(\ref{eq:vorb})) and the wind velocity
$\vec{v}_{\rm{w}} = v_{\rm{w}} \vec{e}_{\rm{R}}$(see, Eq.~(\ref{eq:vw})).
Then, we set a surface which normal vector coincides with $\vec{v}_{\rm{rel}}$.
We consider an accretion region in this surface and estimate the mass and angular momentum transported
to the neighbor of the neutron star.

\begin{figure}
\begin{center}
\includegraphics[width=4cm]{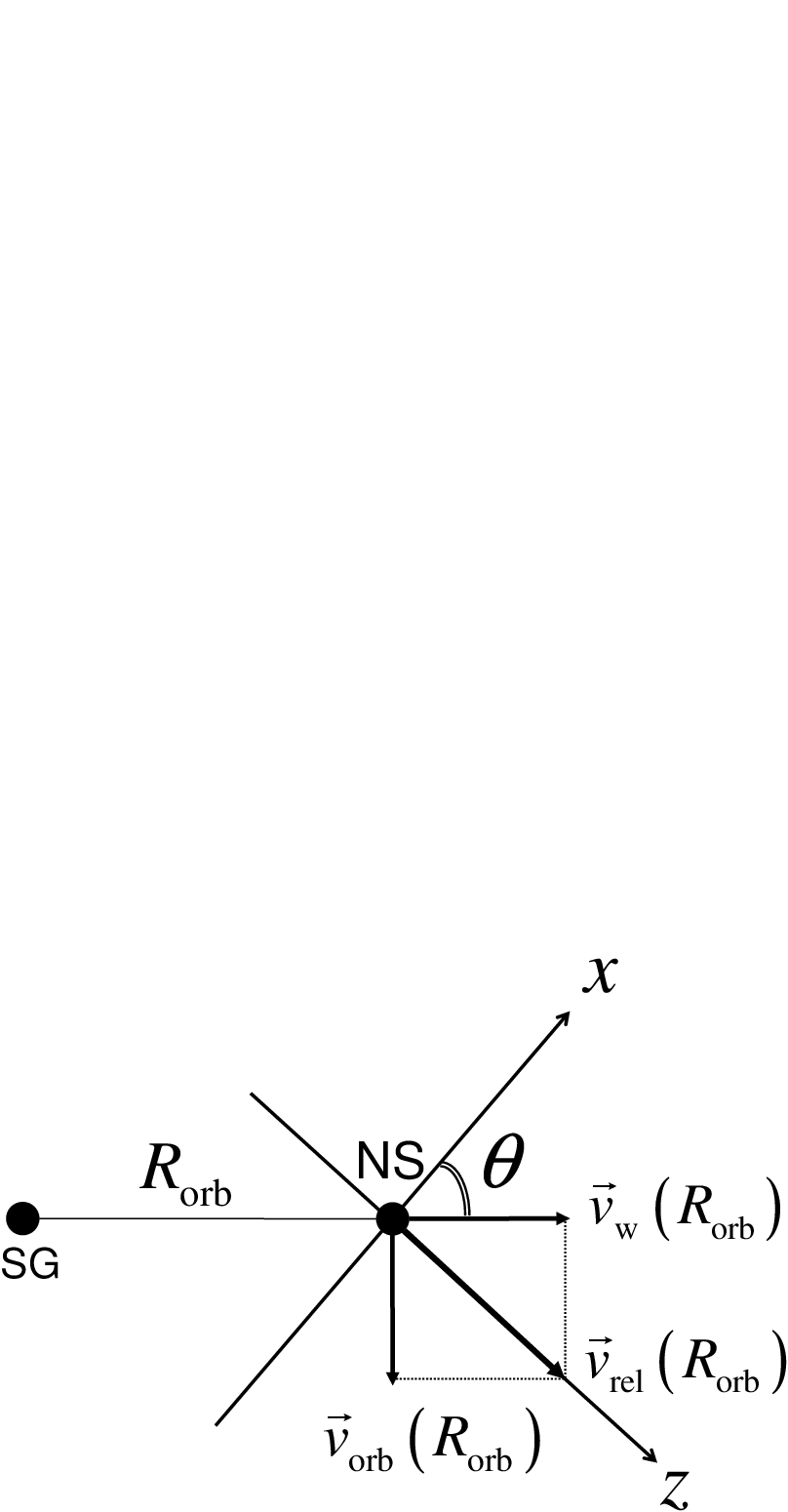}
\end{center}
\caption{
Setting concept of axes. 
The wind, orbital, and relative velocity vectors are those at the neutron star position.
}
\label{fig:angles}
\end{figure}

Now we set $x$-axis on the line of intersection between $S$ plane and the orbital plane.
Then, the $x$-direction is inclined to the relative velocity vector by the angle $\theta$ given as
\begin{equation}
\theta = \frac{\pi}{2} - \tan^{-1} \left(\frac{v_{\rm{w}}}{v_{\rm{orb}}} \right) .
\end{equation}
We set the $z$-direction as the direction of the relative velocity of the wind at the neutron star position.
For $x$- and $z$-axes, we set the positive direction as the direction where recedes away from the donor.
Additionally, we set $y$ direction as $\vec{e}_{\rm{z}} = \vec{e}_{\rm{x}} \times \vec{e}_{\rm{y}}$, where
$\vec{e}_{\rm{x,y,z}}$ denote the unit vectors in each directions.
This situation is depicted in Fig.~\ref{fig:angles}.

\begin{figure}
\begin{center}
\includegraphics[width=4cm]{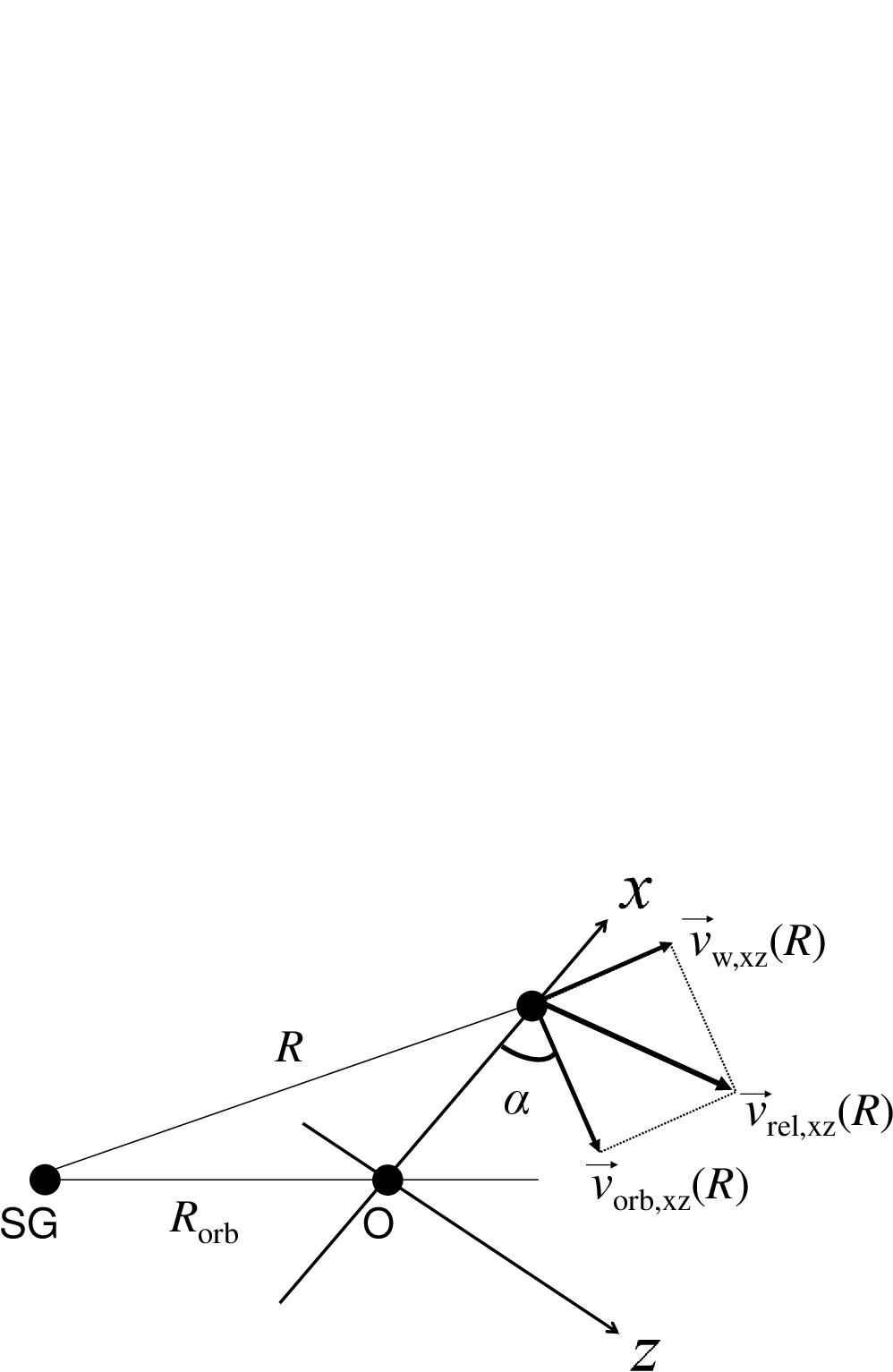}
\end{center}
\caption{Definition of angles.
The wind, orbital, and relative velocity vectors are projected on the $x-z$ plane.
}
\label{fig:angles2}
\end{figure}

On the $S$ plane i$x-y$ planej, the relative velocity of the wind is obtained as
\begin{equation}
v_{\rm{rel}} (x,y) = \left( v_{\rm{w}} (R^{\prime})^2 + v_{\rm{orb}} (R)^2 \right)^{1/2}
\end{equation}
where $R^{\prime}$ is the distance from the center of the donor:
\begin{equation}
R^{\prime} = \sqrt{R(x,0)^2 + y^2} .
\end{equation}
The value of $R$ on the $x$-axis, $R(x)$ could be obtained as
\begin{equation}
R^2 = R_{\rm{orb}}^2 + x^2 - 2 R_{\rm{orb}} x \cos (\pi - \theta) .
\end{equation}
Under such an setting of the coordinate system, we evaluate the balance between wind kinetic energy and
the neutron star potential, and find the accretion region.
In our $x-y$ coordinate system, the potential of the neutron star can be described as
\begin{equation}
U = - \frac{G M_{\rm{NS}}}{\sqrt{x^2 + y^2}} ,
\end{equation}
and the kinetic energy of the wind can be written as
\begin{equation}
K = \frac{1}{2} v_{\rm{rel}}(x,y)^2
\end{equation}
The wind matter passing each point on $S$ surface, $(x,y)$, will be trapped by the neutron star if $K + U < 0$.
Making the dependence on the $x-y$ coordinate clear, Eqs.~(\ref{eq:Mdot}) and (\ref{eq:Jdot}) can be written as
\begin{equation}
\dot{M} = \int_{K+U<0} \rho_{\rm{w}} (x,y) v_{\rm{rel,z}} (x,y) dx dy ,
\label{eq:APMdot}
\end{equation}
and
\begin{equation}
\dot{J} = \int_{K+U<0} \rho_{\rm{w}} (x,y) v_{\rm{rel,z}} (x,y)^2 x dx dy ,
\label{eq:APJdot}
\end{equation}
respectively.
$v_{\rm{rel,z}} (x,y)$ is the $z$-component of the relative velocity vector of the wind at the point $(x,y)$
(namely, the normal component to the $S$ plane), and it is given as the following;
\begin{equation}
v_{\rm{rel,z}} = \frac{ v_{\rm{w}} R}{R^{\prime}}
\sin \left( \frac{\pi}{2} - \alpha \right) + |\vec{v}_{\rm{orb}}| \sin \beta .
\end{equation}
$\alpha$ is the angle between the relative velocity vector of the wind at the position $(x,y)$ and
the reference plane $S$, which is given by
\begin{equation}
\alpha = \frac{\pi}{2} - \cos^{-1} \left( \frac{R^2 + x^2 - R_{\rm{orb}}^2}{2 R x} \right)
\end{equation}
(see Fig.~\ref{fig:angles2}).

\end{document}